\documentclass[aip,apl,reprint]{revtex4-1}
\usepackage{amsmath,amssymb,graphicx,hyperref}
\begin{document}

\title{A Constructor-Theoretic and Quantum Information Approach to the Three-Step Photoemission Model: A Theoretical Investigation}

\author{Saransh Malhotra}
\affiliation{Department of Physics, University of Liverpool}

\begin{abstract}
A novel theoretical reformulation of the conventional three–step photoemission model is presented by integrating the conceptual frameworks of constructor theory and quantum information theory. Each step of the photoemission process—photon absorption, electron transport, and electron emission—is formalized as a physical task (in the constructor–theoretic sense) and modeled by a quantum channel with an explicit operator–level description. This dual framework not only recovers the standard efficiency factorization 
\[
\eta = A\,T\,D,
\]
but also reveals new insights into the preservation of coherence and the interdependence of successive tasks. Furthermore, potential experimental setups and conditions under which the predicted phenomena could be observed are discussed. The motivation of this work is to pave the path for improved implementations of the current photoemission model and to contribute toward the realization of a universal constructor for quantum computation.
\end{abstract}

\maketitle

\section{Introduction}
Photoemission—the process by which electrons are emitted from a material following photon absorption—is central to techniques such as photoelectron spectroscopy and the design of advanced electron sources \cite{Fowler1931,DuBridge1955}. Traditionally, the three–step model decomposes photoemission into: (i) photon absorption and electron excitation, (ii) electron transport from the bulk to the surface, and (iii) electron emission via tunneling through a surface barrier. Although phenomenologically successful, these conventional models do not capture the deep quantum informational structure underlying the process \cite{Hufner2003,Grioni1996}.

Recent developments in \emph{constructor theory} \cite{Deutsch2013,Marletto2015,Marletto2017} have recast physical laws in terms of tasks that are possible or impossible, while \emph{quantum information theory} \cite{NielsenChuang2000,Kraus1983,Watrous2018} describes state evolution via quantum channels—completely positive trace–preserving maps that capture operator–level dynamics \cite{Holevo2012,Preskill1998}. This paper leverages these frameworks to rigorously formalize each photoemission step as a constructor task and to model the associated dynamics by quantum channels, thereby exposing hidden coherence and interference effects.

The motivation of this work is twofold: (i) to pave the way for improved implementations of the current photoemission model by uncovering underlying quantum informational processes, and (ii) to contribute toward the realization of a universal constructor for quantum computation by exploring fundamental task-based implementations in quantum systems \cite{DeutschMarletto2017}.

\section{Theoretical Framework}
\subsection{Constructor Theory and Physical Tasks}
Constructor theory reformulates physics in terms of tasks—sets of input-output pairs describing a transformation on a substrate. For example, the task associated with photon absorption and electron excitation is expressed as:
\[
\mathcal{T}_{\rm abs}: \{\,|g,\text{photon}\rangle \rightarrow |e\rangle\,\}.
\]
A task is considered \emph{possible} if there exists a constructor—an entity capable of reliably performing the transformation repeatedly without degradation—subject to physical constraints such as energy conservation, symmetry, and the no–cloning theorem \cite{Marletto2015,Deutsch2013}. Additional discussions on the formalism and implications of constructor theory can be found in \cite{Marletto2017,DeutschMarletto2017}.

\subsection{Quantum Information Theoretic Representation}
In quantum information theory, the evolution of a quantum system is represented by a quantum channel:
\[
\mathcal{E}: \rho_{\rm in} \rightarrow \rho_{\rm out},
\]
where \(\mathcal{E}\) is a completely positive trace–preserving (CPTP) map \cite{NielsenChuang2000,Kraus1983}. Each step in the photoemission process is modeled as a channel:
\[
\begin{aligned}
\mathcal{E}_1 &: \text{photon absorption \& electron excitation},\\[1mm]
\mathcal{E}_2 &: \text{electron transport},\\[1mm]
\mathcal{E}_3 &: \text{electron emission (tunneling)}.
\end{aligned}
\]
For instance, the absorption process is implemented by an isometric operator \(U_1\) such that
\[
\rho_{\rm exc} = \mathcal{E}_1(\rho_{\rm in}) = U_1 \rho_{\rm in} U_1^\dagger,
\]
with success probability \(A\) \cite{Holevo2012,Preskill1998}. Constructor theory mandates that such operators exist only when all physical constraints are satisfied.

\subsection{Integrated Channel Composition}
The overall photoemission process is modeled as the sequential application of the three quantum channels:
\[
\rho_{\rm out} = \mathcal{E}_3 \circ \mathcal{E}_2 \circ \mathcal{E}_1 (\rho_{\rm in}),
\]
with corresponding success probabilities \(A\), \(T\), and \(D\) for the three steps, respectively. Consequently, the overall efficiency is given by:
\[
\eta = A\,T\,D.
\]
The operator–level descriptions reveal that coherence and interference effects during electron transport and emission may be exploited to optimize \(\eta\) beyond conventional predictions \cite{Watrous2018}.

\section{Mathematical Derivations}
\subsection{Photon Absorption and Electron Excitation}
The interaction Hamiltonian under the dipole approximation is
\[
H_{\rm int} = -e\,\mathbf{r}\cdot \mathbf{E}(\mathbf{r},t),
\]
with the quantized electric field expressed as:
\begin{align}
\mathbf{E}(\mathbf{r},t) &= i\sum_{\mathbf{k},\lambda} \sqrt{\frac{\hbar \omega_k}{2\epsilon_0 V}} \left[ \hat{a}_{\mathbf{k},\lambda}\,\boldsymbol{\epsilon}_{\mathbf{k},\lambda}\,e^{i\mathbf{k}\cdot\mathbf{r}-i\omega_k t} \right. \nonumber\\[1mm]
&\quad \left. - \hat{a}^\dagger_{\mathbf{k},\lambda}\,\boldsymbol{\epsilon}^*_{\mathbf{k},\lambda}\,e^{-i\mathbf{k}\cdot\mathbf{r}+i\omega_k t} \right].
\end{align}
First–order perturbation theory yields the matrix element:
\[
M_{fi} = -e\,i\,\sqrt{\frac{\hbar \omega}{2\epsilon_0 V}}\,\sqrt{n_{\mathbf{k},\lambda}}\,\langle f|\mathbf{r}\cdot \boldsymbol{\epsilon}|i\rangle.
\]
Fermi’s golden rule then provides:
\[
W_{fi} = \frac{2\pi}{\hbar}\,|M_{fi}|^2\,\delta(E_f - E_i - \hbar\omega),
\]
and integration over the density of states \(\rho(E_f)\) gives:
\[
A = \frac{2\pi}{\hbar}\int dE_f\,|M_{fi}|^2\,\delta(E_f - E_i - \hbar\omega)\,\rho(E_f).
\]
In addition to \cite{Fowler1931,DuBridge1955}, the detailed treatment in \cite{CohenTannoudji1977,Feynman1965} provides further background on perturbation methods and transition rates.

\subsection{Electron Transport}
Electron transport to the surface is described by the retarded Green’s function:
\[
\psi(\mathbf{r}_s;E) = \int d^3r_0\,G(\mathbf{r}_s,\mathbf{r}_0;E)\,\psi_0(\mathbf{r}_0).
\]
A conventional approximation is:
\[
G(\mathbf{r}_s,\mathbf{r}_0;E) \propto \exp\!\left(-\frac{|\mathbf{r}_s-\mathbf{r}_0|}{\lambda}\right),
\]
where \(\lambda\) is the inelastic mean free path. In the quantum channel framework, the transport process is modeled as:
\[
\mathcal{E}_2(\rho) = \sum_{k} K_k\,\rho\,K_k^\dagger,
\]
which captures both amplitude attenuation and residual coherence effects. The transport efficiency is expressed as:
\[
T = \int_{0}^{\infty}dz\,\rho_{\rm exc}(z)\,\exp\!\left(-\frac{z}{\lambda}\right).
\]
For further details on the modeling of electron transport, see \cite{AshcroftMermin1976,Callaway1991}. Additional insights into the preservation of coherence in electron transport are discussed in \cite{Perfetti2007,Damascelli2003}.

\subsection{Electron Emission}
Electron emission is modeled by the one–dimensional Schrödinger equation in the barrier region:
\[
-\frac{\hbar^2}{2m}\frac{d^2}{dx^2}\psi(x) + \Phi\,\psi(x) = E\,\psi(x),
\]
where \(\Phi\) is the work function. Using the WKB approximation, the tunneling probability is:
\[
D = \exp\!\left(-2\int_{0}^{d}\sqrt{\frac{2m(\Phi-E)}{\hbar^2}}\,dx\right).
\]
This process is implemented by the channel \(\mathcal{E}_3\), whose operator representation suggests that tailoring the barrier profile may lead to constructive interference effects that enhance \(D\). For a more comprehensive treatment of tunneling and emission processes, refer to \cite{Bardeen1961,Tersoff1985}.

\section{Experimental Considerations}
Although the focus of this work is theoretical, the framework naturally suggests experimental conditions under which the predicted phenomena—particularly enhanced coherence preservation—could be observed:

\begin{itemize}
    \item \textbf{Time-Resolved Photoemission Spectroscopy:}  
    Ultrafast pump-probe experiments, as discussed in \cite{Perfetti2007,Kiss2008}, can resolve the dynamics of electron excitation and transport. Materials with high crystalline quality or low-dimensional structures (e.g., graphene or quantum wells) are promising candidates for observing prolonged coherence times \cite{Damascelli2003}.
    
    \item \textbf{Surface Barrier Engineering:}  
    External electric fields or the deposition of ultra-thin coatings (cf. \cite{Hufner2003}) can be used to tailor the surface work function \(\Phi\). Such modifications may enhance the tunneling probability \(D\) by inducing constructive interference effects at the barrier, which could be detected by changes in the photoemission current \cite{Tersoff1985}.
    
    \item \textbf{Low-Dimensional Systems:}  
    Angle-resolved photoemission spectroscopy (ARPES) studies, as exemplified in \cite{Damascelli2003} and further discussed in \cite{Grioni1996}, on low-dimensional systems like nanowires or layered materials can reveal deviations from the classical exponential decay model in electron transport due to enhanced coherence.
    
    \item \textbf{Temperature and Doping Studies:}  
    Experiments conducted at low temperatures and with controlled doping levels, as indicated in \cite{BreuerPetruccione2002}, can reduce scattering (electron-phonon and electron-electron) and thus preserve coherence. These conditions would help isolate the quantum informational effects predicted by the model.
    
    \item \textbf{Interferometric Techniques:}  
    Interferometry, which directly measures electron phase coherence, may be employed to detect the predicted interference effects within the transport and emission channels \cite{Altshuler1985,Beenakker1997}.
\end{itemize}

These experimental considerations offer concrete pathways for validating the theoretical predictions and for bridging the gap between abstract theory and practical implementations.

\section{Discussion}
The integrated framework reinterprets the conventional three–step photoemission model by merging constructor theory with quantum information theory. Key insights include:
\begin{itemize}
    \item \textbf{Operator-Level Insight:} The quantum channel representation exposes coherence and interference effects during electron transport and emission that are not captured by simple exponential decay models \cite{Watrous2018,Preskill1998}.
    \item \textbf{Task Admissibility:} The constructor–theoretic view rigorously constrains each step to only those transformations that are physically possible, linking energy conservation and symmetry constraints directly to the existence of corresponding quantum operators \cite{Marletto2015,Deutsch2013}.
    \item \textbf{Interdependence of Channels:} The sequential composition \(\mathcal{E}_3 \circ \mathcal{E}_2 \circ \mathcal{E}_1\) reveals that improvements in one channel (e.g., enhanced coherence preservation in \(\mathcal{E}_2\)) may compensate for losses in another, suggesting pathways for overall efficiency optimization \cite{Watrous2018,Beenakker1997}.
\end{itemize}
While the final efficiency remains expressed as \(\eta = A\,T\,D\), the integrated framework provides a deeper understanding of the underlying processes and paves the way for improved implementations of the photoemission model.

\section{Conclusion}
A rigorous theoretical investigation of the three–step photoemission model has been presented by integrating constructor theory and quantum information theory. Each step—photon absorption, electron transport, and electron emission—is formalized as a physical task and modeled as a quantum channel. This dual description not only reproduces the conventional efficiency factorization,
\[
\eta = A\,T\,D,
\]
but also reveals new operator–level effects and interdependencies that suggest pathways for optimizing photoemission efficiency. The motivation of this work is to pave the way for improved implementations of the current model and to contribute toward the realization of a universal constructor for quantum computation. The experimental considerations outlined herein offer a roadmap for future investigations aimed at verifying these theoretical predictions and bridging the gap between abstract theory and practical realization.


\end{document}